\documentclass[aps,pre,twocolumn,groupedaddress]{revtex4}

\usepackage{graphicx}
\usepackage{dcolumn}
\usepackage{bm}
\usepackage{amssymb}

\begin{document}

\title{Entropy-driven aggregation in multilamellar membranes}
\author{Hiroshi Noguchi}
\email[]{noguchi@issp.u-tokyo.ac.jp}
\affiliation{
Institute for Solid State Physics, University of Tokyo,
 Kashiwa, Chiba 277-8581, Japan}

\begin{abstract}
Membrane-fluctuation-induced attraction between ligand--receptor sites binding neighboring membranes
is studied using meshless membrane simulations and the Weil--Farago two-dimensional lattice model.
For the adhesion sites binding two membranes,
this entropic interaction is too weak by itself for the adhesion sites to form a large stable domain.
However, it is found that this attraction is enhanced sufficiently to induce large domains 
either when the sites bind three or more neighboring membranes together or have anchors that harden surrounding membranes. 
The latter effect is understood by the Asakura--Oosawa type of effective potential in the depletion theory.
\end{abstract}

\maketitle

\section{Introduction}

Cell adhesion is a fundamental process required for the removal 
of foreign bodies in immune response, for tissue formation, and for cell motility. 
It is mediated by specific (ligand--receptor) and non-specific interactions. 
Recently, lateral interactions between ligand--receptor adhesion sites in  membranes
have been received growing attention \cite{tana05,smit09,acha10,weik09,kaiz04,smit08,brui94,krob07,fara10,weil10,weil11,spec11,spec12}.
The cooperative aggregation of the adhesion sites yields various patterns of adhesion domains
and also morphological changes in cells and liposomes.

Experimentally, lipid membranes supported on a solid substrate
 are widely used
to study immune reaction and protein functions as well as membrane adhesion \cite{tana05,smit09,acha10}.
Several types of anchoring molecules have been developed 
to control the distance and interactions between the membranes or the membrane and the substrate.
In traditional adhesion experiments using the supported membrane,
the receptors are immobile on the substrate,
while their partners (ligands) are mobile in the fluid membrane.
Recent  experiments with mobile receptors have revealed that
their mobility strengthens the adhesion \cite{smit09,smit08}.
Diffusion of the receptors can induce a high density of the ligand--receptor bonds
in the adhesion domain.

Currently, 
 entropic interactions between (permanently bonded) ligand--receptor
sites  is a topic of active discussion \cite{brui94,krob07,fara10,weil10,weil11,spec11}.
The membrane height fluctuations yield a repulsive force  $f \sim d^{-3}$
between tensionless fluid membranes with a neighboring membrane distance $d$ \cite{helf73}.
Since the adhesion bond holds two membranes close to each other,
an effective attraction works between the adhesion sites.
If the adhesion sites are aggregated,
the rest of the regions of the membranes are allowed to have large height fluctuations.
This entropy gain is the source of this attraction.
Such an interaction can be considered analogous to hydrophobic or depletion interactions, {\it i.e.},
entropy loss of water molecules surrounding
 hydrophobic molecules or polymers surrounding colloids \cite{asak54,lekk11}.
In contrast to these interactions,
the membrane interactions are long-range;
the potential of the mean forces between the membranes and adhesion sites
decays as $\sim r^{-2}$ \cite{fara10}.
Thus, a different type of aggregation behavior can be expected.

Several groups have been investigated this entropic interaction theoretically \cite{brui94,fara10,weil10,weil11,spec11} 
and via simulations \cite{krob07,fara10}.
These studies have reported a weak attraction between adhesion sites, which induces small temporal clusters.
However, it has been concluded that this force is too weak to form a large cluster by itself,
and thus, the researchers have included additional pairwise interactions to investigate the phase separation.
The aim of this letter is to determine the condition required to form a large stable cluster 
only via this membrane-mediated entropic interaction
and to clarify its difference from the typical phase separation generated by a pairwise interaction.
We examine two conditions to intensify the attraction:
(1) increasing entropy of the height fluctuations by binding more than two membranes
and (2) increasing suppression effects on local height fluctuations surrounding the anchoring sites.

We employ one of the solvent-free meshless membrane models \cite{nogu09,nogu06} 
to tackle this problem.
Since we study large-scale membrane fluctuations,
the detailed structures of the bilayer are negligible,
so that the membrane is considered as a curved surface.
To discretize the membrane, 
mesh membrane methods such as the square mesh method used in \cite{krob07}
are also available.
Here, we choose meshless membranes to avoid the influence of the mesh structures on the cluster structures.
In our meshless model, one membrane particle represents a patch of the bilayer membrane
and a membrane can be spontaneously formed.
We also use the  Weil--Farago two-dimensional (2D) lattice model \cite{weil10} 
to understand the membrane-mediated interactions as effective potentials.

\section{Methods}

\subsection{Meshless Membrane Simulation}

\begin{figure}
\includegraphics[width=8.5cm]{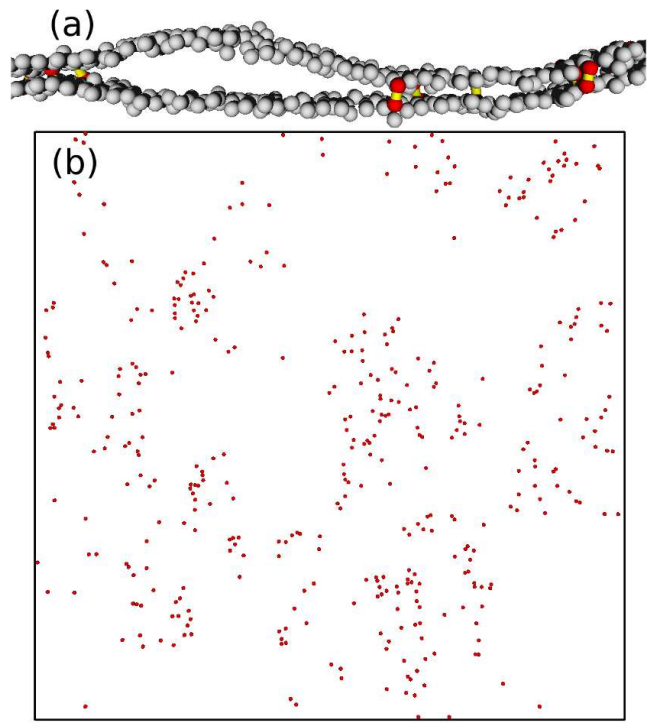}
\caption{Snapshots of fluid membranes
at $N_{\rm {lay}}=2$, $\phi=0.016$, and $\kappa_{\rm r}=1$.
(a) Side view of a sliced snapshot.
The bonded membrane particles and bond connections
are displayed as dark gray (red) spheres
and light gray (yellow) cylinders, respectively.
The other membrane particles are displayed in light gray.
(b) Top view.
The bonded membrane particles in the bottom-layer membrane
 are shown.
The black square frame represents the periodic boundary.
}
\label{fig:snap_l2}
\end{figure}

In this study, we consider $N_{\rm {lay}}$ layers of quasi-planar fluid membranes.
Each membrane is represented by a self-assembled one-layer sheet of $N_{\rm {mb}}$  particles.
These membranes are bound by $N_{\rm {bond}}$ permanently bonded adhesion sites, which are represented
by a linear chain of harmonic bonds using a harmonic potential 
$U_{\rm {bond}}=\sum_{i,j\in {\rm chain}} (k_{\rm {bond}}/2)(r_{i,j} -l_{\rm {bond}})^2$
[see Fig. \ref{fig:snap_l2}(a)].

Since the details of the meshless membrane model are described in Refs. \cite{nogu06,nogu06a,shib11},
we briefly explain the model here.
The particles interact with each other via the potential
$U= \varepsilon( U_{\rm {rep}} +  U_{\rm {att}}) + U_{\rm {\alpha}} + U_{\rm {\rm {bond}}}$,
which consists of a soft-core excluded-volume potential $U_{\rm {rep}}$ with a 
diameter $\sigma$, an attractive potential $U_{\rm {att}}$, a 
curvature potential $U_{\alpha}$, and the bond potential $U_{\rm {\rm {bond}}}$.
The bending rigidity $\kappa$ and the line tension $\Gamma$ of the membrane edge can be separately controlled
by  $U_{\alpha}$ and $U_{\rm {att}}$, respectively. 
In order to ensure that the interactions between neighbor membranes are only a short-range repulsion, 
the membrane particles interact with the particles in different membranes 
only via the repulsive potential $U_{\rm {rep}}$,
while  in each membrane these three membrane potentials ($U_{\rm {rep}}$, $U_{\rm {att}}$, $U_{\rm {\alpha}}$)  
are taken over all contained particles.

The excluded-volume potential is given by
$U_{\rm {rep}}=\sum_{i<j} \exp(-20(r_{i,j}/\sigma-1)+0.126)f_{\rm {cut}}(r_{i,j}/\sigma)$.
The interaction is smoothly cutoff by a $C^{\infty}$ cutoff function \cite{nogu06} 
\begin{equation}
f_{\rm {cut}}(s)=
\exp\bigg[A\Big(1+\frac{1}{(|s|/s_{\rm {cut}})^{12} -1}\Big)\bigg]\theta(s_{\rm {cut}}-s),
\end{equation}
where   $A=1$, $s_{\rm {cut}}=1.2$, and
$\theta(s)$ denotes the unit step function.
The potential $U_{\rm {att}}$ is a function of 
 the local density of particles, 
$\rho_i= \sum_{j} f_{\rm {cut}}(r_{i,j}/\sigma)$,
with the parameters $s_{\rm {half}}=1.8$, $s_{\rm {cut}}=2.1$,
and  $A=\ln(2) \{(s_{\rm {cut}}/s_{\rm {half}})^{12}-1\}$.
Here, $\rho_{i}$ denotes the number of particles in a 
sphere whose radius is approximately 
$r_{\rm {att}} = s_{\rm {half}}\sigma$. The potential $U_{\rm {att}}$ 
is given by
$U_{\rm {att}} = \sum_{i} 0.25\ln[1+\exp\{-4(\rho_i-\rho^*)\}]- C$
where $C= 0.25\ln\{1+\exp(4\rho^*)\}$.
This mutibody potential acts as  a pair potential 
$U_{\rm {att}}\simeq -\rho_i$ with the cutoff at $\rho_i\simeq \rho^*$, 
and it can stabilize
the fluid phase of membranes over a wide range of parameter sets.

The curvature potential is given by 
$U_{\rm {\alpha}}= \sum_i k_{\rm {\alpha},i} \alpha_{\rm {pl}}({\bf r}_{i})$,
where $k_{\rm {\alpha},i}=k_{\rm {\alpha}}^{\rm {bond}}$ and $k_{\rm {\alpha},i}=k_{\rm {\alpha}}^{\rm {norm}}$
for the adhesion sites (membrane particles bonded with neighboring membranes)
and normal (unbonded) membrane particles, respectively.
The shape parameter aplanarity $\alpha_{\rm {pl}}$ is defined as
\begin{equation}
\alpha_{\rm {pl}}
= \frac{9\lambda_1\lambda_2\lambda_3}{(\lambda_1+\lambda_2+\lambda_3)
    (\lambda_1\lambda_2+\lambda_2\lambda_3+\lambda_3\lambda_1)},
\end{equation}
where $\lambda_1$, $\lambda_2$, and $\lambda_3$ are
the eigenvalues of the weighted gyration tensor,
$a_{\alpha\beta}= \sum_j (\alpha_{j}-\alpha_{\rm G})
(\beta_{j}-\beta_{\rm G})w_{\rm {mls}}(r_{i,j})$,
where $\alpha, \beta=x,y,z$ and ${\bf r}_{\rm G}=\sum_j {\bf r}_{j}w_{\rm {mls}}(r_{i,j})/\sum_j w_{\rm {mls}}(r_{i,j})$.
The aplanarity $\alpha_{\rm {pl}}$ 
represents the degree of deviation from a plane, and it is proportional to $\lambda_1$
 for $\lambda_1 \ll \lambda_2, \lambda_3$.
A Gaussian function with $C^{\infty}$ cutoff~\cite{nogu06} 
is employed as a weight function 
\begin{equation}
w_{\rm {mls}}(r_{i,j})=
\exp \Big[\frac{(r_{i,j}/1.5\sigma)^2}{(r_{i,j}/3\sigma)^{12} -1}\Big]\theta(r_{\rm {cc}}-r_{i,j}).
\end{equation}

In this letter, we use  $N_{\rm {mb}}=25,600$, $\varepsilon/k_{\rm B}T=4$, $\rho^*=6$, $k_{\rm {bond}}/k_{\rm B}T=40$, 
$l_{\rm {bond}}=1.5\sigma$, and $k_{\rm {\alpha}}^{\rm {norm}}/k_{\rm B}T=5$,
where $k_{\rm B}T$ denotes the thermal energy.
Thus, the tensionless membranes have a bending rigidity $\kappa/k_{\rm B}T= 9$, line tension 
$\Gamma\sigma/k_{\rm B}T=4$ and area $a_0=1.4\sigma^2$ per membrane particle \cite{shib11}.
The nearest-neighbor distance of the membrane particles $l_0=1.1\sigma$ ($\sim$ the membrane thickness of $5$ nm)
is taken as the length unit.
The membranes are simulated in a tensionless state 
in the $N\gamma T$ ensemble (constant
number of particles $N$, surface tension $\gamma$, and temperature $T$)
with a periodic boundary box of $L_x=L_y$
and a Langevin thermostat \cite{fell95,nogu12}.
The numerical errors are estimated from $8$--$128$ independent runs.
To investigate the clustering in the membranes in three or more layers ($N_{\rm {lay}}\geq 3$),
the whole region of membranes is considered to be have one value of $\kappa$,
{\it i.e.}, $\kappa_{\rm r}\equiv k_{\rm {\alpha}}^{\rm {bond}}/k_{\rm {\alpha}}^{\rm {norm}}=1$.
To investigate  a local interaction between the anchoring sites and surrounding lipids,
 the ratio $\kappa_{\rm r}$ is varied for double membranes ($N_{\rm {lay}}=2$). 
The anchoring sites harden the surrounding membranes at $\kappa_{\rm r}>1$.

\subsection{2D Lattice Model}

The potential of the mean force between two adhesion sites with distance $r$
in tensionless membranes is 
derived by  Farago as $\Phi(r) = 2k_{\rm B}\ln(r/l)$,
where $l$ denotes the membrane thickness \cite{fara10}. 
However,
the sum of this potential interaction between all pairs of the adhesion sites
yields too strong an attraction,
since the pair interaction can be shielded by other sites.
Weil and Farago proposed a 2D lattice model
to take into account the multibody nature of the interaction \cite{weil10}.
Since local membrane height fluctuations are mainly suppressed by the nearest adhesion site,
 the entropy of the membrane segment $i$ 
can be expressed by a function of the distance $d_i^{\rm {min}}$ to the nearest site.
The entropy of the height fluctuations are expressed as an effective potential,
\begin{eqnarray}
\label{eq:2dmodel_0}
U_{\rm {2d}} &=& \sum_{i=1}^{N_{\rm {2d}}} u_i(1-s_i) \\
u_i &=& \frac{(N_{\rm {lay}}-1)k_{\rm B}T}{\pi}\Big(\frac{l_0}{d_i^{\rm {min}}}\Big)^2,
\label{eq:2dmodel_1}
\end{eqnarray}
for a triangular lattice with the unit lattice length $l_0$,
and  $N_{\rm {2d}}$ denotes the number of lattice sites.
For the adhesion sites and the unbonded sites
$s_i=1$ and $s_i=0$, respectively, so that
the integration is essentially taken over the unbonded lattice sites.
In order to treat multilamellar membranes,
we slightly extend the  Weil--Farago model by adding a factor $N_{\rm {lay}}-1$ in Eq. (\ref{eq:2dmodel_1}).
Thus, the height differences between neighboring membranes
are assumed to independently fluctuate.
We use the Metropolis Monte Carlo method to obtain the equilibrium states
for the membranes with an almost-square shape of $L_x=149l_0$ and $L_y=86\sqrt{3}l_0$
with $N_{\rm {2d}}=25628$.

\begin{figure}
\includegraphics[width=8.5cm]{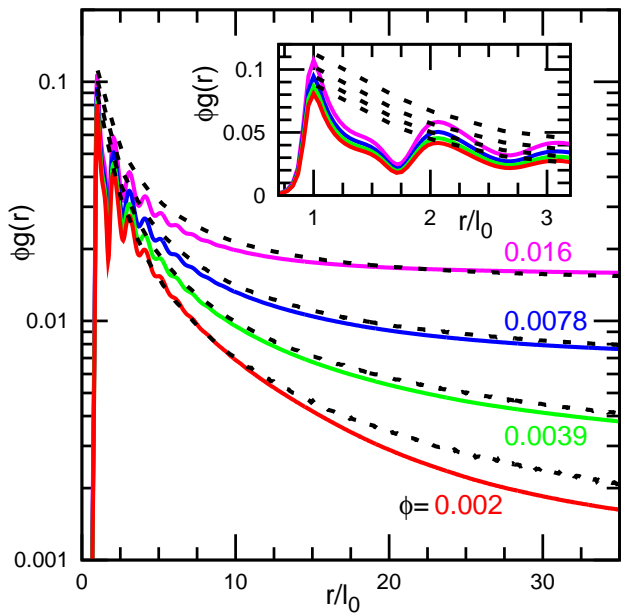}
\caption{
Radial distribution function $g(r)$ between bonded membrane particles
at $N_{\rm {lay}}=2$.
The solid and dashed lines represent the data 
for the meshless membrane simulation with  $\kappa_{\rm r}=1$
and  the 2D lattice model (Eq. (\ref{eq:2dmodel_1})),
respectively.
The inset shows the magnification of the first and second peaks.
}
\label{fig:rdf}
\end{figure}

\section{Double Membranes}

First, we investigate interactions between the adhesion sites binding two membranes ($N_{\rm {lay}}=2$),
and we confirm the conclusions of previous studies \cite{krob07,fara10,weil10}.
The adhesion sites are distributed throughout the membranes 
and their small clusters are temporally formed but do not grow into large stable clusters
(see Fig. \ref{fig:snap_l2}).
Even if a simulation is started from a large cluster,
it gradually dissolves into the mixed state.

The cluster size is almost independent of the mean density of adhesion sites $\phi=N_{\rm {bond}}/N_{\rm {mb}}$.
The radial distribution function $g(r)$ multiplied by $\phi$ for the distance $r$ projected in the $xy$ plane
is shown in Fig. \ref{fig:rdf}.
When all of the particles at distance $r$ are adhesion sites
and the projected membrane density is uniform, $\phi g(r)=1$.
With increasing $\phi$,
the number of contacted particle pairs increases by only $10$\%,
while the density at $r/l_0 \gg 1$ linearly increases.
Thus, the clusters do not grow, and instead 
the excess amount of adhesion sites dissolve in isolation
or form other small clusters.
The pair interactions of adhesion sites are shielded by other surrounding sites,
and the local density in the clusters is saturated.

The Weil--Farago 2D lattice model \cite{weil10} reproduces our simulation results  well
(compare the solid and dashed lines in Fig. \ref{fig:rdf}).
In particular, the heights of the first peak of the simulation and the model coincide.
In the lattice model, $g(r)$ exhibits slightly slower decays,
which are likely caused by the difference in the unit area: $a_0=1.1{l_0}^2$ 
and $a_0=(\sqrt{3}/2){l_0}^2$ for the meshless  and  lattice models, respectively.
If a two-body potential $\Phi(r_{ij})$ is instead employed,
 all adhesion sites assemble into one cluster at $N_{\rm {bond}}\ge 4$.
Thus, the multibodiness of the interactions is crucial to understand this  clustering.

\begin{figure}
\includegraphics[width=8.5cm]{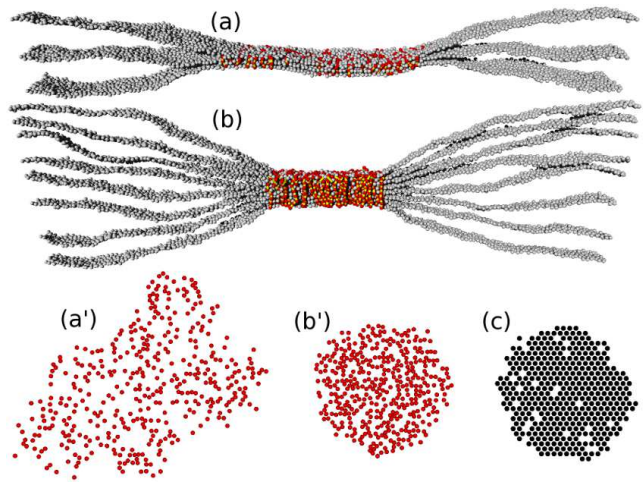}
\caption{
Snapshots of membranes at $\phi=0.016$ and $\kappa_{\rm r}=1$
for (a), (a$'$)  $N_{\rm {lay}}=3$ and (b), (b$'$)  $N_{\rm {lay}}=8$.
(a), (b) Side view of sliced snapshots.
(a$'$), (b$'$) Top view of bonded membrane particles in the middle layer.
(c) Snapshot of the 2D lattice model (Eq. (\ref{eq:2dmodel_1}))  at $\phi=0.016$ and  $N_{\rm {lay}}=8$.
}
\label{fig:snap_l8}
\end{figure}

\begin{figure}
\includegraphics[width=8.5cm]{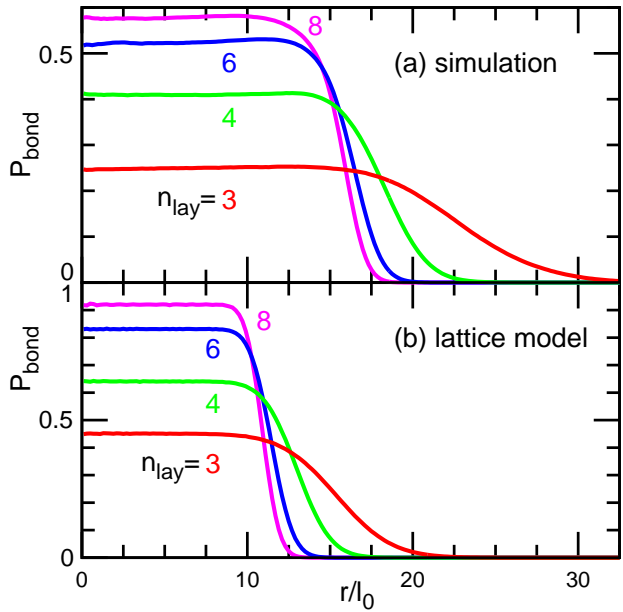}
\caption{
Probability distribution $P_{\rm {bond}}$ of the bonded membrane particles
as a function of the distance from the center of the cluster for $N_{\rm {lay}}=3$, $4$, $6$, and $8$
at $\phi=0.016$ 
calculated by (a) the meshless membrane simulation with  $\kappa_{\rm r}=1$
and (b) the 2D lattice model (Eq. (\ref{eq:2dmodel_1})).
}
\label{fig:rhis0}
\end{figure}

\section{Three or More Membranes}

In order to produce a large stable cluster,
the bending entropy of membranes is enhanced by the addition of more layers of membranes.
For triple membranes ($N_{\rm {lay}}=3$),
the adhesion sites form a single large domain,
whose shape shows large fluctuations [see Figs. \ref{fig:snap_l8}(a) and (a$'$)].
A few sites often leave the domain but soon return before moving far away,
since an isolated site will further suppress the membrane fluctuations of the larger area.
With increasing $N_{\rm {lay}}$,
the domain becomes more compact and circular  [see Figs. \ref{fig:snap_l8} (b) and (b$'$)].

The distribution of the number ratio $P_{\rm {bond}}$ of bonded membrane particles 
in the middle layer of the membranes is shown in Fig. \ref{fig:rhis0}(a).
The number ratio is uniform in the middle of the domain.
Interestingly, many unbonded membrane particles still remain in the domain
even at $N_{\rm {lay}}=8$ ($P_{\rm {bond}}=0.58$).

\begin{figure}
\includegraphics[width=8.5cm]{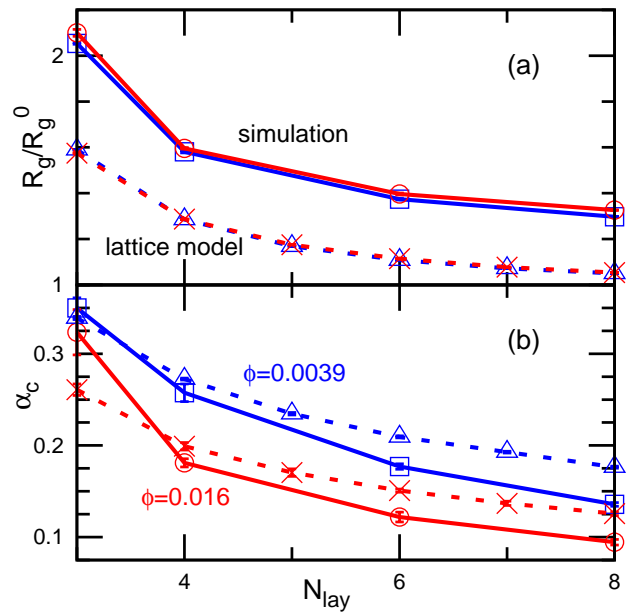}
\caption{
The $N_{\rm {lay}}$ dependence of
(a) radius of gyration $R_{\rm g}$ and (b) shape parameter $\alpha_{\rm c}$
of the adhesion sites
for $\phi=0.0039$ ($\Box, \triangle$) and $\phi=0.016$ ($\circ, \times$).
The solid and dashed lines represent the data for the meshless membrane simulation with  $\kappa_{\rm r}=1$
and the 2D lattice model (Eq. (\ref{eq:2dmodel_1})), respectively.
}
\label{fig:rg_l}
\end{figure}

The $N_{\rm {lay}}$ dependences of two shape parameters are shown in Fig. \ref{fig:rg_l}.
The radius of gyration $R_{\rm g}$ is normalized by $R_{\rm g}^{0}=\sqrt{a_0N_{\rm {bond}}/2\pi}$ 
for a densely packed circular domain.
The shape deviation from a circular disk is calculated as 
$\alpha_{\rm c}=(\nu_1-\nu_2)/(\nu_1+\nu_2)$,
where $\nu_1$ and $\nu_2$ denote two eigenvalues of the gyration tensor
of the adhesion sites.
Although the domain shape becomes circular,
 its size approaches not $R_{\rm g}/R_{\rm g}^0= 1$ but $1.3$.
This saturation is caused by the existence of the unbonded particles in the domain.

The domain shapes at $\phi=0.0039$ and $\phi=0.016$ ($N_{\rm {bond}}=100$, $400$) 
are also compared in Fig. \ref{fig:rg_l}.
The normalized size $R_{\rm g}/R_{\rm g}^0$ coincides very well.
Smaller domains have slightly larger values of $\alpha$,
since the same amplitude of the boundary fluctuation yields larger effects on the whole shape
for a smaller domain.
At $\phi=0.0039$,
 $P_{\rm {bond}}$ also exhibits a distribution similar to that at $\phi=0.016$
(data not shown).
Thus, the domain formation is not sensitive to the domain size,
at least when the domain is sufficiently smaller than the membrane area.

A single domain is formed when $N_{\rm {lay}}\ge 3$ also in the 2D lattice model
(see Figs. \ref{fig:snap_l8}-\ref{fig:rg_l}).
The 2D domains are more compact and they comprise a lesser number of unbonded sites
than the domains in the meshless membrane simulations.
In the lipid bilayer membranes,
the height fluctuations
are governed by molecular protrusions
smaller in length than the membrane thickness \cite{goet99}.
This protrusion is taken into account in the meshless membrane model
but not in the lattice model.
The absence of the protrusion likely causes the reduction in the domain size.
Except for this domain size difference,
the lattice model reproduces the domain properties very well.
Thus, the addition of the membrane layers can be simply
interpreted as a linear entropy increase of the membrane height fluctuations.

\section{Double Membranes with Membrane-Hardening Anchors}

Membrane proteins often modify the surrounding membranes.
In particular, the effects of height mismatch between the hydrophobic cores of the proteins
and the hydrophobic tails of lipids have been well investigated \cite{deme08,west09}.
Here, we simply consider the effect of the anchor proteins 
on the main quantity being examined in this study, {\it i.e.},  bending rigidity $\kappa$.
The anchors of ligands or receptors suppress the bending fluctuations of 
the surrounding membranes.

\begin{figure}
\includegraphics[width=8.5cm]{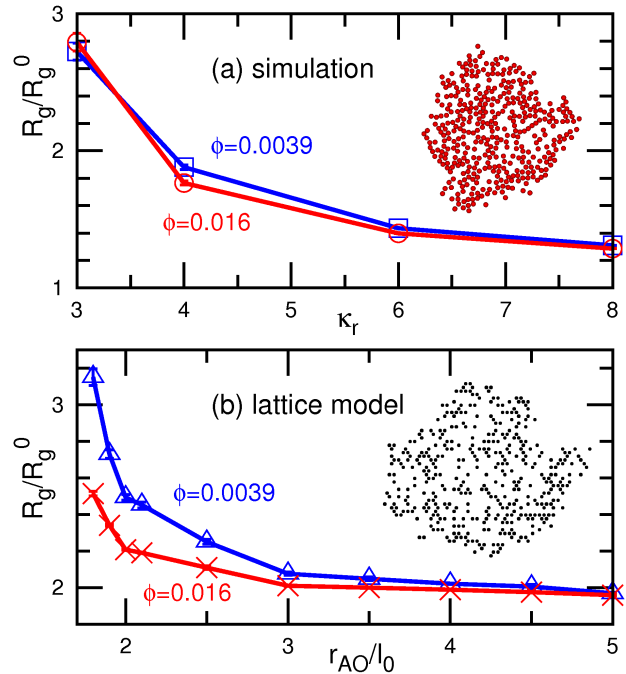}
\caption{
Variation in the radius of gyration $R_{\rm g}$ of the adhesion sites with the membrane-hardening anchors
at  $N_{\rm {lay}}=2$ for $\phi=0.0039$ ($\Box, \triangle$) and $\phi=0.016$ ($\circ, \times$).
(a) Dependence on the ratio of  bending rigidity $\kappa_{\rm r}$ in the meshless membrane simulation.
(b) Dependence on the the depletion radius $r_{\rm {AO}}$ in the 2D depletion lattice model
(Eq. (\ref{eq:2dmodel_ao})).
Snapshots at $\kappa_{\rm r}=8$ and  $r_{\rm {AO}}=5l_0$ are shown in the insets of (a) and (b), respectively.
}
\label{fig:rg_r}
\end{figure}

In the meshless membrane model, 
the locally large bending rigidity is given by
a large value of $k_{\rm {\alpha},i}=k_{\rm {\alpha}}^{\rm {bond}}$ 
of the bonded membrane particles.
In the limit $\kappa_{\rm r}=k_{\rm {\alpha}}^{\rm {bond}}/k_{\rm {\alpha}}^{\rm {norm}}\to \infty$,
the neighboring membrane at a distance $r<3\sigma$ from the adhesion sites
becomes completely flat.
This local flattening induces a stable domain formation
even at $N_{\rm {lay}}=2$.
With increasing $\kappa_{\rm r}$,
the domain radius decreases [see Fig. \ref{fig:rg_r}(a)].

In order to take this effect into account,
the potential, given by Eq. (\ref{eq:2dmodel_1}), 
in the 2D lattice model
 is modified as
\begin{equation}
\label{eq:2dmodel_ao}
u_i=\left\{ 
\begin{array}{ll}
\frac{k_{\rm B}T}{\pi}\Big(\frac{l_0}{d_i^{\rm {min}}-r_{\rm {AO}}}\Big)^2
& ( d_i^{\rm {min}} \ge r_{\rm {AO}}+l_0) \\
\frac{k_{\rm B}T}{\pi}
& ( d_i^{\rm {min}} < r_{\rm {AO}}+l_0). 
\end{array}
\right.
\end{equation}
The surrounding membrane sites at the distance $r<r_{\rm {AO}}$
are flat and have no bending fluctuations.
This treatment is similar to the Asakura--Oosaka theory for the depletion interaction \cite{asak54,lekk11}.

As the height fluctuations of several neighbor sites are suppressed at $r_{\rm {AO}}/l_0 \gtrsim 2$,
a single domain is formed [see Fig. \ref{fig:rg_r}(b)].
At large values of $r_{\rm {AO}}$,
the domain radius is saturated to $R_{\rm g}/R_{\rm g}^0= 2$,
which is considerably larger than the value $R_{\rm g}/R_{\rm g}^0= 1.3$
of the meshless simulations.
Thus, the omission of the protrusions
works differently from the case of the $N_{\rm {lay}}$ increase.
It reduces the effective attractions.
Since the protrusion is also suppressed at large values of $\kappa_{\rm r}$,
the surrounding unbonded membrane particles lose more entropy
than in the case of the 2D lattice representation.

\section{Summary and Discussions}

We have revealed that the clustering of the adhesion sites binding neighboring membranes
can be caused only by  entropic interactions via membrane height fluctuations.
The reduction in the membrane fluctuations due to close contact yields
an attractive interaction between the adhesion sites.
For binding between two membranes, this interaction is too weak to form a large stable domain
and increasing the number of the adhesion sites does not lead to cluster growth.
In order to overcome this problem, we extend the system to enhance this attraction in two ways:
(1) The number of membrane layers is increased; 
when three or more membranes are bound,
the adhesion sites form one large domain. 
(2) Adhesion anchors that harden surrounding membranes are employed.
This induces the depletion type of attraction between the adhesion sites,
since the areas of the surrounding membranes are overlapped in the clusters.
Both conditions intensify the entropy gain by the cluster formation
and their effects can be explained by the extended 2D lattice models.

Although ligand--receptor pairs bind only two membranes in living cells,
 ligand--receptor chains connecting several membranes
are synthetically producible.
Our predictions on clustering in multilamellar membranes
can be experimentally examined using supporting membranes.
On the other hand, the latter depletion-like interactions
may play a role in biomembranes and
this type of domain formation can be induced
not only in adhesion anchor proteins but also in other membrane-associated proteins.
It is reported that the adhesion sites are accumulated on fixed membrane boundary \cite{weil11}.
The adhesion sites and other proteins can likely form a domain at a specific region in a plasma membrane.

The domain formations in bound membranes are  different from those of typical phase separation by a pairwise interaction.
When two types of molecules are phase-separated in a binary fluid,
a small fraction of either type of molecules dissolves in the other phase,
and their fractions rapidly decrease with decreasing temperature.
In the case of the membrane adhesion sites,
the competition between the entropies in the perpendicular direction (height fluctuations)
and in the horizontal directions (mixing of the adhesion sites)
determines the phase behavior, so that
it does not  directly depend on the temperature 
(it may indirectly be affected by change in the membrane properties).
While a large amount of unbonded membrane particles dissolves in circular domains,
no adhesion sites dissolve in the membranes even from largely deformed domains.
This asymmetry is caused by the long-range height correlation of the membranes.
Such stable domains involving other lipids and proteins
may form a good platform for biological functions.

\acknowledgments
The author would like to thank
 G. Gompper and H. Shiba for informative discussions.
The numerical calculations were partly
carried out on SGI Altix ICE 8400EX 
at ISSP Supercomputer Center, University of Tokyo. 
This work is supported by KAKENHI (25400425) from
the Ministry of Education, Culture, Sports, Science, and Technology of Japan.


\begin{thebibliography}{25}
\expandafter\ifx\csname natexlab\endcsname\relax\def\natexlab#1{#1}\fi
\expandafter\ifx\csname bibnamefont\endcsname\relax
  \def\bibnamefont#1{#1}\fi
\expandafter\ifx\csname bibfnamefont\endcsname\relax
  \def\bibfnamefont#1{#1}\fi
\expandafter\ifx\csname citenamefont\endcsname\relax
  \def\citenamefont#1{#1}\fi
\expandafter\ifx\csname url\endcsname\relax
  \def\url#1{\texttt{#1}}\fi
\expandafter\ifx\csname urlprefix\endcsname\relax\def\urlprefix{URL }\fi
\providecommand{\bibinfo}[2]{#2}
\providecommand{\eprint}[2][]{\url{#2}}

\bibitem[{\citenamefont{Tanaka and Sackmann}(2005)}]{tana05}
\bibinfo{author}{\bibfnamefont{M.}~\bibnamefont{Tanaka}} \bibnamefont{and}
  \bibinfo{author}{\bibfnamefont{E.}~\bibnamefont{Sackmann}},
  \bibinfo{journal}{Nature} \textbf{\bibinfo{volume}{437}},
  \bibinfo{pages}{656} (\bibinfo{year}{2005}).

\bibitem[{\citenamefont{Smith and Sackmann}(2009)}]{smit09}
\bibinfo{author}{\bibfnamefont{A.-S.} \bibnamefont{Smith}} \bibnamefont{and}
  \bibinfo{author}{\bibfnamefont{E.}~\bibnamefont{Sackmann}},
  \bibinfo{journal}{ChemPhysChem} \textbf{\bibinfo{volume}{10}},
  \bibinfo{pages}{66} (\bibinfo{year}{2009}).

\bibitem[{\citenamefont{Achalkumar et~al.}(2010)\citenamefont{Achalkumar,
  Bushby, and Evans}}]{acha10}
\bibinfo{author}{\bibfnamefont{A.~S.} \bibnamefont{Achalkumar}},
  \bibinfo{author}{\bibfnamefont{R.~J.} \bibnamefont{Bushby}},
  \bibnamefont{and} \bibinfo{author}{\bibfnamefont{S.~D.} \bibnamefont{Evans}},
  \bibinfo{journal}{Soft Matter} \textbf{\bibinfo{volume}{6}},
  \bibinfo{pages}{6036} (\bibinfo{year}{2010}).

\bibitem[{\citenamefont{Weikl et~al.}(2009)\citenamefont{Weikl, Asfaw, Heinrich
  Krobath~and, and Lipowsky}}]{weik09}
\bibinfo{author}{\bibfnamefont{T.~R.} \bibnamefont{Weikl}},
  \bibinfo{author}{\bibfnamefont{M.}~\bibnamefont{Asfaw}},
  \bibinfo{author}{\bibfnamefont{B.~R.} \bibnamefont{Heinrich Krobath~and}},
  \bibnamefont{and} \bibinfo{author}{\bibfnamefont{R.}~\bibnamefont{Lipowsky}},
  \bibinfo{journal}{Soft Matter} \textbf{\bibinfo{volume}{5}},
  \bibinfo{pages}{3213} (\bibinfo{year}{2009}).

\bibitem[{\citenamefont{Kaizuka and Groves}(2004)}]{kaiz04}
\bibinfo{author}{\bibfnamefont{Y.}~\bibnamefont{Kaizuka}} \bibnamefont{and}
  \bibinfo{author}{\bibfnamefont{J.~T.} \bibnamefont{Groves}},
  \bibinfo{journal}{Biophys. J.} \textbf{\bibinfo{volume}{86}},
  \bibinfo{pages}{905} (\bibinfo{year}{2004}).

\bibitem[{\citenamefont{Smith et~al.}(2008)\citenamefont{Smith, Sengupta,
  Goennenwein, Seifert, and Sackmann}}]{smit08}
\bibinfo{author}{\bibfnamefont{A.-S.} \bibnamefont{Smith}},
  \bibinfo{author}{\bibfnamefont{K.}~\bibnamefont{Sengupta}},
  \bibinfo{author}{\bibfnamefont{S.}~\bibnamefont{Goennenwein}},
  \bibinfo{author}{\bibfnamefont{U.}~\bibnamefont{Seifert}}, \bibnamefont{and}
  \bibinfo{author}{\bibfnamefont{E.}~\bibnamefont{Sackmann}},
  \bibinfo{journal}{Proc.\ Natl.\ Acad.\ Sci.\ USA}
  \textbf{\bibinfo{volume}{105}}, \bibinfo{pages}{6906} (\bibinfo{year}{2008}).

\bibitem[{\citenamefont{Bruinsma et~al.}(1994)\citenamefont{Bruinsma, Goulian,
  and Pincus}}]{brui94}
\bibinfo{author}{\bibfnamefont{R.}~\bibnamefont{Bruinsma}},
  \bibinfo{author}{\bibfnamefont{M.}~\bibnamefont{Goulian}}, \bibnamefont{and}
  \bibinfo{author}{\bibfnamefont{P.}~\bibnamefont{Pincus}},
  \bibinfo{journal}{Biophys. J.} \textbf{\bibinfo{volume}{67}},
  \bibinfo{pages}{746} (\bibinfo{year}{1994}).

\bibitem[{\citenamefont{Krobath et~al.}(2007)\citenamefont{Krobath, Sch{\"u}tz,
  Lipowsky, and Weikl}}]{krob07}
\bibinfo{author}{\bibfnamefont{H.}~\bibnamefont{Krobath}},
  \bibinfo{author}{\bibfnamefont{G.~J.} \bibnamefont{Sch{\"u}tz}},
  \bibinfo{author}{\bibfnamefont{R.}~\bibnamefont{Lipowsky}}, \bibnamefont{and}
  \bibinfo{author}{\bibfnamefont{T.~R.} \bibnamefont{Weikl}},
  \bibinfo{journal}{EPL} \textbf{\bibinfo{volume}{78}}, \bibinfo{pages}{38003}
  (\bibinfo{year}{2007}).

\bibitem[{\citenamefont{Farago}(2010)}]{fara10}
\bibinfo{author}{\bibfnamefont{O.}~\bibnamefont{Farago}},
  \bibinfo{journal}{Phys. Rev. E} \textbf{\bibinfo{volume}{81}},
  \bibinfo{pages}{050902} (\bibinfo{year}{2010}).

\bibitem[{\citenamefont{Weil and Farago}(2010)}]{weil10}
\bibinfo{author}{\bibfnamefont{N.}~\bibnamefont{Weil}} \bibnamefont{and}
  \bibinfo{author}{\bibfnamefont{O.}~\bibnamefont{Farago}},
  \bibinfo{journal}{Euro. Phys. J. E} \textbf{\bibinfo{volume}{33}},
  \bibinfo{pages}{81} (\bibinfo{year}{2010}).

\bibitem[{\citenamefont{Weil and Farago}(2011)}]{weil11}
\bibinfo{author}{\bibfnamefont{N.}~\bibnamefont{Weil}} \bibnamefont{and}
  \bibinfo{author}{\bibfnamefont{O.}~\bibnamefont{Farago}},
  \bibinfo{journal}{Phys. Rev. E} \textbf{\bibinfo{volume}{84}},
  \bibinfo{pages}{051907} (\bibinfo{year}{2011}).

\bibitem[{\citenamefont{Speck}(2011)}]{spec11}
\bibinfo{author}{\bibfnamefont{T.}~\bibnamefont{Speck}},
  \bibinfo{journal}{Phys. Rev. E} \textbf{\bibinfo{volume}{83}},
  \bibinfo{pages}{050901} (\bibinfo{year}{2011}).

\bibitem[{\citenamefont{Speck and Vink}(2012)}]{spec12}
\bibinfo{author}{\bibfnamefont{T.}~\bibnamefont{Speck}} \bibnamefont{and}
  \bibinfo{author}{\bibfnamefont{R.~L.~C.} \bibnamefont{Vink}},
  \bibinfo{journal}{Phys. Rev. E} \textbf{\bibinfo{volume}{86}},
  \bibinfo{pages}{031923} (\bibinfo{year}{2012}).

\bibitem[{\citenamefont{Helfrich}(1973)}]{helf73}
\bibinfo{author}{\bibfnamefont{W.}~\bibnamefont{Helfrich}},
  \bibinfo{journal}{Z.\ Naturforsch} \textbf{\bibinfo{volume}{28c}},
  \bibinfo{pages}{693} (\bibinfo{year}{1973}).

\bibitem[{\citenamefont{Asakura and Oosawa}(1954)}]{asak54}
\bibinfo{author}{\bibfnamefont{S.}~\bibnamefont{Asakura}} \bibnamefont{and}
  \bibinfo{author}{\bibfnamefont{F.}~\bibnamefont{Oosawa}},
  \bibinfo{journal}{J. Chem. Phys.} \textbf{\bibinfo{volume}{22}},
  \bibinfo{pages}{1255} (\bibinfo{year}{1954}).

\bibitem[{\citenamefont{Lekkerkerker and Tuinier}(2011)}]{lekk11}
\bibinfo{author}{\bibfnamefont{H.~N.~W.} \bibnamefont{Lekkerkerker}}
  \bibnamefont{and} \bibinfo{author}{\bibfnamefont{R.}~\bibnamefont{Tuinier}},
  \emph{\bibinfo{title}{Colloids and the Depletion Interaction}}
  (\bibinfo{publisher}{Springer}, \bibinfo{address}{Dordrecht},
  \bibinfo{year}{2011}).

\bibitem[{\citenamefont{Noguchi}(2009)}]{nogu09}
\bibinfo{author}{\bibfnamefont{H.}~\bibnamefont{Noguchi}},
  \bibinfo{journal}{J.\ Phys.\ Soc.\ Jpn.} \textbf{\bibinfo{volume}{78}},
  \bibinfo{pages}{041007} (\bibinfo{year}{2009}).

\bibitem[{\citenamefont{Noguchi and Gompper}(2006{\natexlab{a}})}]{nogu06}
\bibinfo{author}{\bibfnamefont{H.}~\bibnamefont{Noguchi}} \bibnamefont{and}
  \bibinfo{author}{\bibfnamefont{G.}~\bibnamefont{Gompper}},
  \bibinfo{journal}{Phys.\ Rev.\ E} \textbf{\bibinfo{volume}{73}},
  \bibinfo{pages}{021903} (\bibinfo{year}{2006}{\natexlab{a}}).

\bibitem[{\citenamefont{Noguchi and Gompper}(2006{\natexlab{b}})}]{nogu06a}
\bibinfo{author}{\bibfnamefont{H.}~\bibnamefont{Noguchi}} \bibnamefont{and}
  \bibinfo{author}{\bibfnamefont{G.}~\bibnamefont{Gompper}},
  \bibinfo{journal}{J.\ Chem.\ Phys.} \textbf{\bibinfo{volume}{125}},
  \bibinfo{pages}{164908} (\bibinfo{year}{2006}{\natexlab{b}}).

\bibitem[{\citenamefont{Shiba and Noguchi}(2011)}]{shib11}
\bibinfo{author}{\bibfnamefont{H.}~\bibnamefont{Shiba}} \bibnamefont{and}
  \bibinfo{author}{\bibfnamefont{H.}~\bibnamefont{Noguchi}},
  \bibinfo{journal}{Phys. Rev. E} \textbf{\bibinfo{volume}{84}},
  \bibinfo{pages}{031926} (\bibinfo{year}{2011}).

\bibitem[{\citenamefont{Feller et~al.}(1995)\citenamefont{Feller, Zhang,
  Pastor, and Brooks}}]{fell95}
\bibinfo{author}{\bibfnamefont{S.~E.} \bibnamefont{Feller}},
  \bibinfo{author}{\bibfnamefont{Y.}~\bibnamefont{Zhang}},
  \bibinfo{author}{\bibfnamefont{R.~W.} \bibnamefont{Pastor}},
  \bibnamefont{and} \bibinfo{author}{\bibfnamefont{B.~R.}
  \bibnamefont{Brooks}}, \bibinfo{journal}{J.\ Chem.\ Phys.}
  \textbf{\bibinfo{volume}{103}}, \bibinfo{pages}{4613} (\bibinfo{year}{1995}).

\bibitem[{\citenamefont{Noguchi}(2012)}]{nogu12}
\bibinfo{author}{\bibfnamefont{H.}~\bibnamefont{Noguchi}},
  \bibinfo{journal}{Soft Matter} \textbf{\bibinfo{volume}{8}},
  \bibinfo{pages}{3146} (\bibinfo{year}{2012}).

\bibitem[{\citenamefont{Goetz et~al.}(1999)\citenamefont{Goetz, Gompper, and
  Lipowsky}}]{goet99}
\bibinfo{author}{\bibfnamefont{R.}~\bibnamefont{Goetz}},
  \bibinfo{author}{\bibfnamefont{G.}~\bibnamefont{Gompper}}, \bibnamefont{and}
  \bibinfo{author}{\bibfnamefont{R.}~\bibnamefont{Lipowsky}},
  \bibinfo{journal}{Phys.\ Rev.\ Lett.} \textbf{\bibinfo{volume}{82}},
  \bibinfo{pages}{221} (\bibinfo{year}{1999}).

\bibitem[{\citenamefont{{de Meyer} et~al.}(2008)\citenamefont{{de Meyer},
  Venturoli, and Smit}}]{deme08}
\bibinfo{author}{\bibfnamefont{F.~J.} \bibnamefont{{de Meyer}}},
  \bibinfo{author}{\bibfnamefont{M.}~\bibnamefont{Venturoli}},
  \bibnamefont{and} \bibinfo{author}{\bibfnamefont{B.}~\bibnamefont{Smit}},
  \bibinfo{journal}{Biophys.\ J.} \textbf{\bibinfo{volume}{95}},
  \bibinfo{pages}{1851} (\bibinfo{year}{2008}).

\bibitem[{\citenamefont{West et~al.}(2009)\citenamefont{West, Brown, and
  Schmid}}]{west09}
\bibinfo{author}{\bibfnamefont{B.}~\bibnamefont{West}},
  \bibinfo{author}{\bibfnamefont{F.~L.~H.} \bibnamefont{Brown}},
  \bibnamefont{and} \bibinfo{author}{\bibfnamefont{F.}~\bibnamefont{Schmid}},
  \bibinfo{journal}{Biophys.\ J.} \textbf{\bibinfo{volume}{96}},
  \bibinfo{pages}{101} (\bibinfo{year}{2009}).

\end{thebibliography}
\end{document}